\documentclass[prb, twocolumn, showpacs, superscriptaddress, preprintnumbers, floatfix]{revtex4-1}  

\usepackage{graphicx}
\usepackage{color}

\draft 

\begin{document}


\title{Photoluminescence spectroscopy of pure pentacene, perfluoropentacene and mixed thin films}



\author{F.~Anger}
\affiliation{Institut f\"ur Angewandte Physik, Universit\"at T\"ubingen, 72076 T\"ubingen, Germany}

\author{J. O. Oss\'o}
\affiliation{MATGAS 2000 AIE, Campus de la UAB, 08193 Bellaterra, Spain}

\author{U.~Heinemeyer}
\affiliation{Institut f\"ur Angewandte Physik, Universit\"at T\"ubingen, 72076 T\"ubingen, Germany}

\author{K.~Broch}
\affiliation{Institut f\"ur Angewandte Physik, Universit\"at T\"ubingen, 72076 T\"ubingen, Germany}

\author{R.~Scholz}
\affiliation{Institut f\"ur Angewandte Photophysik, TU Dresden, 01069 Dresden, Germany}

\author{A.~Gerlach}
\affiliation{Institut f\"ur Angewandte Physik, Universit\"at T\"ubingen, 72076 T\"ubingen, Germany}

\author{F.~Schreiber}
\email[]{frank.schreiber@uni-tuebingen.de}
\affiliation{Institut f\"ur Angewandte Physik, Universit\"at T\"ubingen, 72076 T\"ubingen, Germany}



\date{\today}

\begin{abstract}
  We report detailed temperature dependent photoluminescence (PL) spectra of pentacene (PEN), perfluoropentacene (PFP), and PEN:PFP mixed thin films grown on SiO$_2$.
  PEN and PFP are particularly suitable for this study, since they are structurally compatible for good intermixing and form a model donor/acceptor system.
  The PL spectra of PEN are discussed in the context of existing literature and compared to the new findings for PFP.
  We analyze the optical transitions observed in the spectra of PEN and PFP using time-dependent density functional theory (TD-DFT) calculations.
  Importantly, for the mixed PEN:PFP film we observe an optical transition in PL at 1.4~eV providing evidence for coupling effects in the blend. 
  We discuss a possible charge-transfer (CT) and provide a tentative scheme of the optical transitions in the blended films.
\end{abstract}

\maketitle 


\section{Introduction} \label{Sec:1_Introduction}

There has been growing interest in organic semiconducting materials in the last decade. 
A major driving force behind this is obviously their potential for electronic and optoelectronic applications, many of which actually involve several organic components, frequently as donor/acceptor systems.~\cite{Book_Bruetting_2005,Wagner_2010_AFM_20_4295,Book_Salaneck_2001,Witte_2004_JoMR_19_1889,Koch_2007_C_8_1438}
The coupling between these compounds including the energy level alignment and possible charge transfer (CT) states are not well understood, but they are crucially important, since they determine to a large extent the resulting electronic and optical properties including absorption and emission cross sections.
One of the fundamental properties is photoluminescence (PL), which is also an important way of testing the interactions and the effective energy levels~\cite{Salzmann_2008_JotACS_130_12870,Aoki-Matsumoto_2001_IjompB_15_3753,He_2005_AApl_87_211117,Park_2002_APL_80_2872}.
In order to maximize the potential for coupling effects, good intermixing of the respective two compounds is desirable, for which structural compatibility is beneficial.
Pentacene (PEN, C$_{22}$H$_{14}$) and perfluoropentacene (PFP, C$_{22}$F$_{14}$)~\cite{Sakamoto_2006_MCaLC_444_225,Fujii_2009_JoESaRP_174_65,Inoue_2005_JJoAP_44_3663,Hinderhofer_2007_Tjocp_127_194705,Broch_2011_PRB_83_245307,Salzmann_2008_JotACS_130_12870,Medina_2007_JoCP_126_111101,Koch_2008_JotACS_130_7300} are promising candidates for a model system, since on the one hand they exhibit obviously different electron affinities and may thus act as a donor/acceptor pair, and on the other hand they are expected to be structurally compatible due their similar molecular geometry.
Moreover, they exhibit high charge carrier mobilities and have already been tested in device structures.~\cite{Anthony_2008_ACIE_47_452,Inoue_2005_JJoAP_44_3663,Hinderhofer_2007_Tjocp_127_194705}
In fact, detailed X-ray experiments on PEN:PFP mixed thin films~\cite{Salzmann_2008_L_24_7294,Hinderhofer_2011_JoCP_134_104702} have shown good intermixing of the molecules promoting the probability of molecular interaction between PEN and PFP. 
While an equimolar mixed thin film leads to PEN:PFP intermixed phases, non-1:1 mixing ratios result in phase separation between homogeneously blended and pure phases.~\cite{Hinderhofer_2011_JoCP_134_104702}
In this study, using temperature dependent PL spectroscopy on PEN:PFP mixed films, we find evidence for coupling between PEN and PFP.
In particular, we discover a new optical transition in PL at 1.4~eV that is related to an absorption band at 1.6~eV~\cite{Broch_2011_PRB_83_245307}.

\section{Experimental} \label{Sec:2_Experimental}

Pure PEN (Sigma Aldrich, 99.9~\% purity) and PFP (Kanto Denka Kogyo Co., 99~\% purity) as well as coevaporated blended thin films with mixing ratio 4:1, 1:1, and 1:2 (PEN:PFP) were grown on oxidized silicon wafers using organic molecular beam deposition~\cite{Witte_2004_JoMR_19_1889,Schreiber_2004_JoPCM_16_881} techniques.
Thin films were grown under ultra high vacuum (UHV) conditions with a thickness of roughly 20~nm on substrates kept at 325~K. 
In this regime PEN thin films nucleate in the 'thin film phase'~\cite{Mattheus_2003_JotACS_125_6323}, whereas for PFP so far only one phase has been reported~\cite{Sakamoto_2006_MCaLC_444_225}.
Further details on the growth and structure, including X-ray diffraction data can be found in Ref.~\citenum{Hinderhofer_2011_JoCP_134_104702}.
In addition, X-ray data of the blended PEN:PFP films shown in the supplementary material are essentially identical to the ones presented in Ref.~\citenum{Hinderhofer_2011_JoCP_134_104702} and hence demonstrate good intermixing.
The mixing ratio of the blends was determined by X-ray photoelectron spectroscopy on similar samples.~\cite{Broch_2011_PRB_83_245307}

Low temperature photoluminescence spectra were acquired using a Horiba Jobin Yvon LabRam HR 800 spectrometer with a CCD-1024$\times$256-FIVS-3S9.
The samples were mounted on a KONTI cryostat and cooled down by liquid nitrogen/helium under high vacuum (HV) conditions. 
Measurements were performed during warm up.
The average time interval between consecutive spectra in a series was more than one hour.

Excitation was performed using an Ar$^+$-laser with lines at 488~nm (2.54~eV) and 514~nm (2.41~eV). 
Spectra obtained at the different excitation energies yield essentially identical shapes. 
While the more intense line at 2.41~eV leads to a higher quantum yield of the spectra, the line at 2.54~eV covers a broader spectral range.
Thus, we present temperature dependent PL of the pure samples excited at 488~nm, since features above 2~eV for PEN become more detailed.
For the PL of the mixtures, which is rather redshifted with respect to the PL of the pure samples, we make use of the more intense line at 514~nm. 
The diameter of the laser spot on the sample was approximately 1~$\mu$m, hence the spectra represent an average over phase separated PEN, PFP, and mixed islands as they occur for non-1:1 mixing ratios~\cite{Hinderhofer_2011_JoCP_134_104702}.

The luminescence spectra were corrected for instrumental sensitivity. 
Normalization was performed with respect to the dominant Raman peak of the Si-substrate at 520~cm$^{-1}$. 
In order to avoid photo-oxidation effects affecting in particular the blends, the samples were stored and measured under HV conditions.

\section{Pure materials} \label{Sec:3_Results}

In this section we present the PL spectra of the pure materials and discuss them in the light of time-dependent density funtional theory (TD-DFT) calculations.

\subsection{Pentacene (PEN)}

\begin{figure}
  \includegraphics[width=8cm]{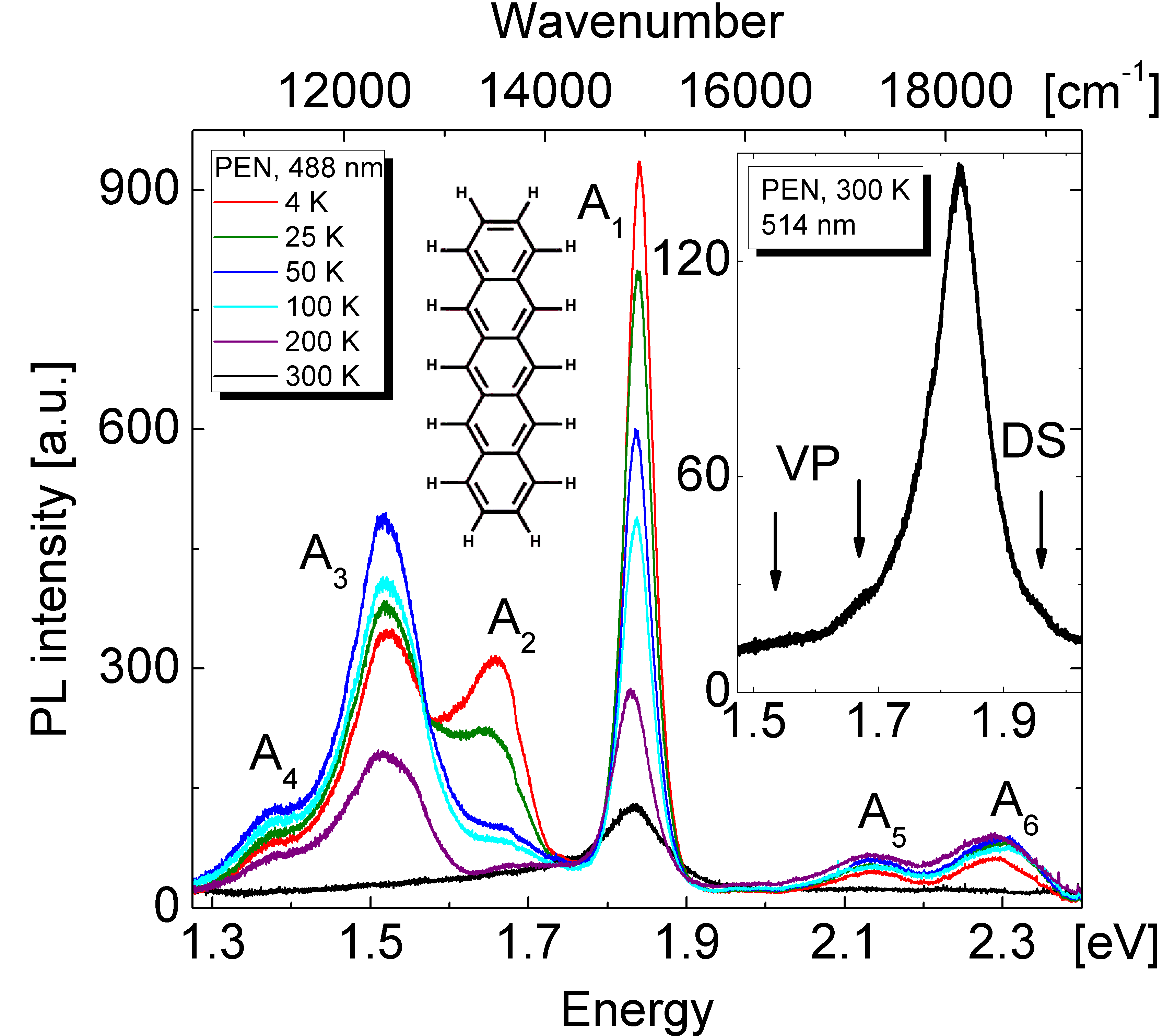}
  \caption{\label{PEN} PL spectra of a polycrystalline PEN thin film (20~nm) at different temperatures, excited at 488~nm (2.54~eV). While at RT, there is only one band ($A_1$) at 1.83~eV observable, other bands ($A_2$-$A_6$) arise towards lower temperatures. At RT (inset, PL excited at 514~nm), we observe a weak vibronic progression (VP) and a shoulder denoted as Davydov-splitting (DS)~\cite{He_2005_AApl_87_103107}. For further details, see text.}
\end{figure}

In order to relate to existing results in the literature~\cite{Aoki-Matsumoto_2001_IjompB_15_3753,He_2005_AApl_87_211117,Park_2002_APL_80_2872,He_2010_APL_96_263303,He_2005_AApl_87_103107}, we first discuss pure PEN films.
Fig.~\ref{PEN} shows the PL of PEN at different temperatures. 
We can identify six PL bands $A_1$ to $A_6$.
The dominating PL line $A_1$ lies at 1.83~eV, which is close to the peak energy of the lowest absorption band at 1.85~eV reported for a thin film with similar properties.~\cite{Hinderhofer_2007_Tjocp_127_194705} 
It is the only band visible at room temperature (RT) and exhibits two broad shoulders shifted by $\sim$150~meV towards its lower energy side (see inset). 
We observe another shoulder blueshifted by $\sim$100~meV from the maximum. 
The intensity of peak $A_1$ increases strongly towards lower temperatures, whereas the blueshifted shoulder and the redshifted shoulders keep their low intensity. 
The latter become superimposed by other bands at 1.52~eV and 1.66~eV. Towards 4~K, the peak $A_1$ is subject to a blueshift by approximately 10~meV.

Below 250~K further peaks arise at 1.52~eV ($A_3$) and 1.37~eV ($A_4$), respectively.
In good agreement with Ref.~\citenum{Aoki-Matsumoto_2001_IjompB_15_3753}, we observe a maximum for bands $A_3$ and $A_4$ at approximately 50~K. 
Both peaks grow in parallel exhibiting a constant intensity ratio.
Another band arises at 1.66~eV ($A_2$) for temperatures below 100~K. 
Below 200~K we observe the bands $A_5$ (2.14~eV) and $A_6$ (2.29~eV).
Unlike other PL features, these bands do not change notably in intensity towards lower temperatures.

$A_1$ is known in the literature as the free exciton band~\cite{Aoki-Matsumoto_2001_IjompB_15_3753,He_2005_AApl_87_211117}, originating from the $S_1 \rightarrow S_0$ transition.
Its position varies between 1.76~eV in thicker thin films~\cite{Park_2002_APL_80_2872} and 1.88~eV in monolayers with a parabolically increasing intensity for small film thicknesses~\cite{He_2010_APL_96_263303}.
At 1.82~eV it has been observed less pronounced also in PEN crystals~\cite{Aoki-Matsumoto_2001_IjompB_15_3753}.
The shoulder 100~meV above $A_1$ coincides with previous reports, where in particular for very thin PEN films and clusters a higher Davydov component of band $A_1$ was observed.~\cite{He_2005_AApl_87_103107}

We assign the two shoulders of peak $A_1$ denoted as VP in the inset of Fig.~\ref{PEN} to a vibronic progression.
Interestingly, these shoulders do not become more pronounced at lower temperatures.
A similar behavior of the vibronic progression was found in tetracene which also crystallizes in a herringbone structure, where this effect could be explained by superradiance.~\cite{Lim_2004_PRL_92_107402}

$A_2$ was assigned to a self-trapped exciton (STE) state.
Its strongly pronounced signal increasing towards low temperature is reported in particular for PEN crystals.~\cite{Aoki-Matsumoto_2001_IjompB_15_3753,He_2005_AApl_87_211117} 
The origin of peak $A_3$ remains unclear.
It has been discussed as a deep STE~\cite{Aoki-Matsumoto_2001_IjompB_15_3753} or charge-transfer exciton~\cite{Knupfer_2006_CP_325_92}. 
Also a transition related to impurities of PEN crystals was found in this region.~\cite{He_2005_AApl_87_211117}

\subsection{Perfluoropentacene (PFP)}

\begin{figure}
  \includegraphics[width=8cm]{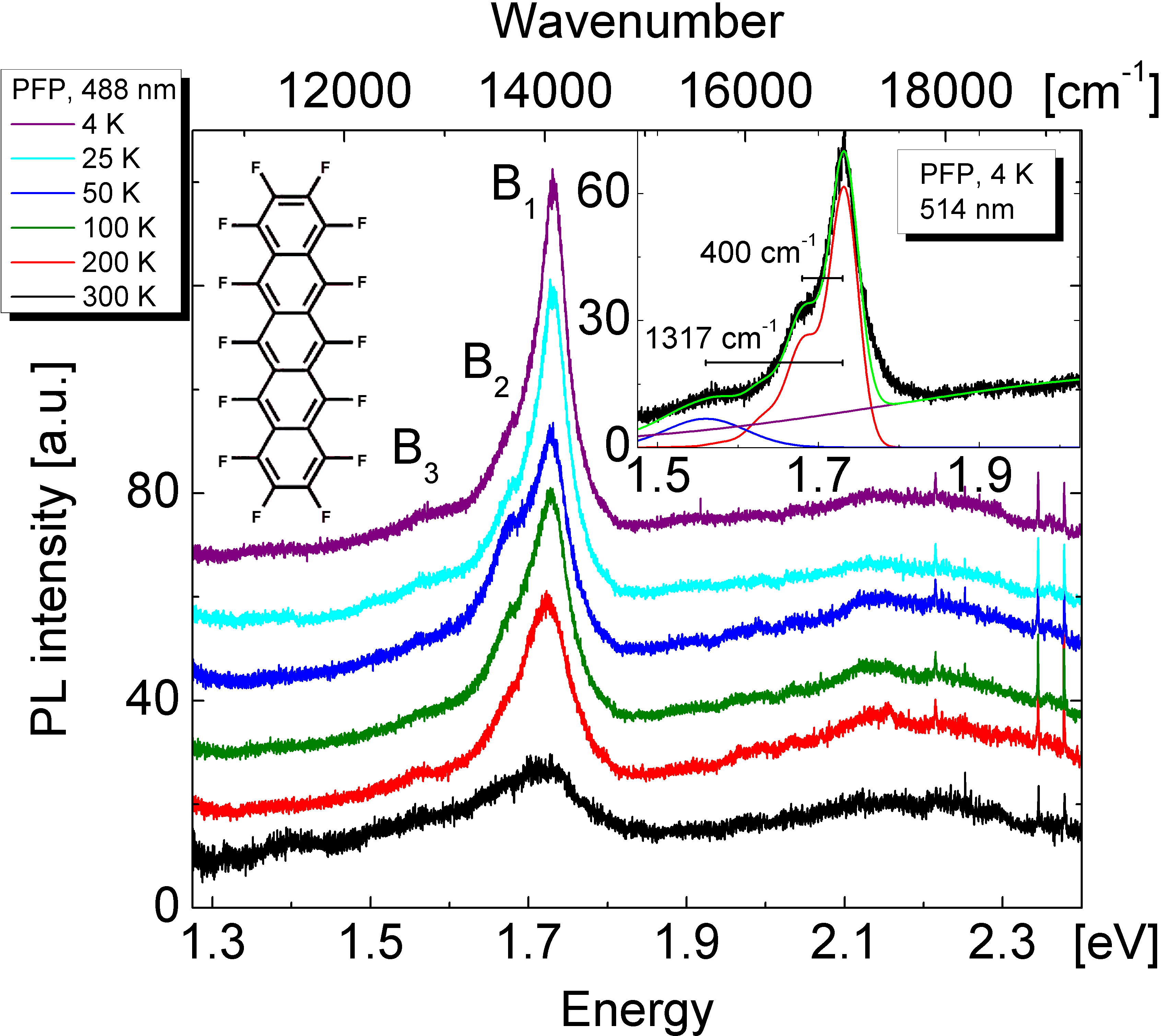}
  \caption{\label{PFP} PL spectra of PFP at varying temperature show a band $B_1$ at 1.72~eV. The spectra are excited at 488~nm (2.54~eV) and offset for clarity. Redshifted from $B_1$, we observe two shoulders $B_2$ and $B_3$, that we tentatively assign to a vibronic progression. The inset (PL excited at 514~nm) shows a model fit of the vibronic progression.}
\end{figure}

PL spectra of pure PFP have not been reported before.
The temperature dependent PL spectrum obtained from a 20~nm thick PFP thin film is presented in Fig.~\ref{PFP}. 
The strong peak at 1.72~eV ($B_1$) exhibits a fine structure towards its low energy tail. 
At lower temperatures the peak becomes more narrow and increases in amplitude. 
We also observe a blueshift of the peak maximum of $B_1$ towards 4~K by $\sim$10~meV. 
The fine structure consists of sidebands at 1.66~eV ($B_2$) and 1.56~eV ($B_3$), which appear as shoulders at RT, but are clearly distinguishable at low temperatures where they become narrower.

We note that the lowest peak in the absorption spectra for a comparable thin film was observed at 1.78~eV~\cite{Hinderhofer_2007_Tjocp_127_194705}, which results in a Stokes-shift of the PL maximum of 60~meV.
Because of the rather small Stokes-shift of peak $B_1$ with respect to the lowest absorption maximum~\cite{Hinderhofer_2007_Tjocp_127_194705}, we assign this peak to the $S_1 \rightarrow S_0$ transition of PFP. 
The fine structure seems to originate from a vibronic progression resulting from elongated internal vibrational modes.
In order to describe the data at 4~K, we perform a model fit of the intensity $I$ using a series of Gaussians that are Poisson-distributed in intensity following
\[ I(E) = A \sum_n (S^n/n!)(1/w_n \sqrt{2 \pi}) exp(-(E-E_n)^2/2 w_n^2) \]
(see inset of~Fig.~\ref{PFP}, red curve), where $E_n$ and $w_n$ denote the energy position and width of the peaks, respectively. We find a Huang-Rhys-factor~$S$ of 0.5 for peak $B_2$.
We complete this model by introducing another Gaussian at 1.56~eV ($B_3$) and a broad peak representing the background (blue and purple curve in the inset, respectively).
The background is formed by weak and broadened peaks energetically above $B_1$, which, however, cannot be resolved properly.
While the red curve corresponds to C-C bending modes at 48~meV, $B_3$ may be related to a vibronic progression of C-C stretching vibrations at 163~meV.~\cite{Fujii_2009_JoESaRP_174_65} 
Tab.~\ref{tab:PEN_PFP} gives a list of selected PEN and PFP peaks.

\begin{table}[t]
  \centering
  \begin{tabular}{ll|l}
    \hline
    \hline
    \multicolumn{3}{c}{PEN} \\
    \hline
       $A_1$ & 1.83-1.84~eV & $S_1 \rightarrow S_0$\\
       $A_2$ & 1.66~eV& Self trapped exciton\\
       $A_3$ & 1.52~eV& Being discussed\\
       $A_4$ & 1.37~eV& Vibronic progression of $A_3$\\
       $A_5$ & 2.14~eV& Recombination of electron-hole pairs\\
       $A_6$ & 2.29~eV& Recombination of electron-hole pairs\\
    \hline
    \multicolumn{3}{c}{PFP} \\
    \hline
       $B_1$ & 1.72~eV & $S_1 \rightarrow S_0$\\
       $B_2$ & 1.66~eV& Vibronic progression of $B_1$   \\
       $B_3$ & 1.56~eV& Vibronic progression of $B_1$   \\
    \hline
    \hline
  \end{tabular}
  \caption{\label{tab:PEN_PFP} Summary of selected peaks observed for PEN and PFP and their assignment. 
  For further details, see text.}
\end{table}

\subsection{Electronic transitions of pure materials. \label{Sec:3c_Transitions}}

The observed absorption \cite{Broch_2011_PRB_83_245307} and PL features can be compared to electronic excitations calculated with TD-DFT \cite{Jamorski_1996_JCP_104_5134} using the hybrid functional B3LYP \cite{Becke_1988_PRA_38_3098,Lee_1988_PRB_37_785} together with a 6-31G(d) variational basis, as implemented in the {\sc gaussian03} program package \cite{gaussian03}, compare Tab.~\ref{table:b3lyp}.

When applied to a free molecule, TD-DFT is known to reproduce the oscillator strength of different electronic transitions reasonably well, but the calculated vertical transition energies $E_{\rm vert}$ in the ground state geometry of the molecule are subject to a rather large variation of about 0.4 eV around the measured transition energies~\cite{Grimme_2003_C_3_292, Hellweg_2008_PCCP_10_4119}. 
In cases where the observed absorption line shape shows a pronounced vibronic progression, it becomes mandatory to distinguish between the observed origin band $E_{00}$ and the average over the observed vibronic transitions, $E_{\rm vert}$. The internal deformations in the relaxed excited geometries of PEN and PFP elongate several breathing modes observable in resonant Raman spectroscopy, and in absorption they result in a pronounced vibronic progression with about $E_{\rm vert}-E_{00}\approx 0.1$~eV \cite{Phdthesis_Gisslen_2009}. 
TD-DFT for a single molecule does not account for the gas-to-crystal shift arising from the surrounding molecules, and lattice sums over long range interactions between the molecular transition dipoles are ignored. 
Nevertheless, together with the calculated oscillator strengths, TD-DFT can provide a one-to-one correspondence between observed and calculated absorption bands.
    
\begin{table*}[htbp]
	\begin{tabular}{cccc|ccc|ccc}
		\hline
		\hline
		\multicolumn{4}{c|}{}                                                    & \multicolumn{3}{c|}{PEN} & \multicolumn{3}{c}{PFP}    \\  
		\cline{5-10}
		State               &   & Configuration                  & Transition Dipole & E            & E  & $f_{\rm osc}$  & E    & E     & $f_{\rm osc}$ \\ 
                    &   &                                &                   & eV           & eV    &   1     & eV   & eV    &  1           \\
                    &   &                                &                   & exp.         & calc. & calc.   & exp. & calc. & calc.        \\ 	
		\hline
		$S_0$               &   & $1^1A_{g}$                      &                   &              & 0      &        &  & 0    &            \\
		$T_1$               &   & $1^3B_{2u}$                     &                   & 0.85$^a$      & 0.99   &        &  & 0.88 &            \\
		\hline
		$S_1 \leftarrow S_0$ &   & $1^1B_{2u} \leftarrow 1^1A_{g}$ & $B_{2u}(y)$        & 2.31$^b$      & 1.94   & 0.041  &  1.99$^c$ & 1.77 &  0.045  \\
                     &   &                               &                   & 2.13$^c$      &        &        &  1.78$^d$ &      &         \\
                     &   &                               &                   & 1.85$^d$      &        &        &           &      &         \\
		$S_2 \leftarrow S_0$ &   & $1^1B_{3g} \leftarrow 1^1A_{g}$ &                   &               & 2.96   &        &           & 2.70 &         \\
		$S_3 \leftarrow S_0$ &   & $1^1B_{1u} \leftarrow 1^1A_{g}$ & $B_{1u}(x)$        & 3.73$^e$      & 3.24   & 0.012  &  2.71$^c$ & 3.01 &  0.337  \\
		$S_4 \leftarrow S_0$ &   & $2^1B_{3g} \leftarrow 1^1A_{g}$ &                   &               & 3.27   &        &           & 3.10 &         \\
		$S_5 \leftarrow S_0$ &   & $2^1B_{2u} \leftarrow 1^1A_{g}$ & $B_{2u}(y)$        &               & 4.01   & 0.000  &           & 3.68 &  0.001  \\
		$S_6 \leftarrow S_0$ & (PEN)  & $1^1B_{1u} \leftarrow 1^1A_{g}$ & $B_{1u} (x)$        & 4.40$^e$      & 4.34   & 3.24   &           &      &         \\ $S_6 \leftarrow S_0$      & (PFP)  & $2^1A_{g} \leftarrow 1^1A_{g}$ &      &               &       &         &           & 4.03 &         \\ 
		\hline
		$T_2 \leftarrow T_1$ &   & $1^3B_{3g} \leftarrow 1^3B_{2u}$ & $B_{1u}(x)$       & 1.2 - 1.4$^f$ & 1.16   & 0.002  &           & 1.11 &  0.004  \\
		$T_3 \leftarrow T_1$ &   & $1^3B_{1u} \leftarrow 1^3B_{2u}$ &                  &               & 2.02   &        &           & 1.69 &         \\
		$T_4 \leftarrow T_1$ &   & $2^3B_{3g} \leftarrow 1^3B_{2u}$ & $B_{1u}(x)$       & 2.46$^g$      & 2.26   & 0.695  &           & 2.16 &  0.581  \\                      &   &                               &                   & 1.97$^h$      &        &        &           &      &         \\
		\hline 
		\hline
	\end{tabular}
	\caption{Comparison between calculated energies of lowest triplet configuration (DFT, B3LYP/6-31G(d)) and electronic transitions (TD-DFT, B3LYP/6-31G(d)) with observed values. 
	For dipole-allowed transitions, the orientation of the transition dipole is indicated ($x$, $B_{1u}(x)$: long axis, $y$, $B_{2u}(y)$: short axis) \\ 
	$^a$ in polycrystalline material\cite{Vilar_1983_CPL_94_522}{, $^b$} free molecule \cite{Amirav_1981_JoPC_85_309}{, $^c$} in dichlorobenzene solution \cite{Hinderhofer_2007_Tjocp_127_194705}{, $^d$} in polycrystalline film \cite{Hinderhofer_2007_Tjocp_127_194705}{, $^e$} in neon matrix \cite{Halasinski_2000_JoPCA_104_7484}{, $^f$} in polycrystalline film \cite{Thorsmoelle_2009_PRL_102_17401}{, $^g$} in cyclohexane solution \cite{Hellner_1972_JCSFT2_68_1928}{, $^h$} in polycrystalline film \cite{Jundt_1995_CPL_241_84}.}
	\label{table:b3lyp}
\end{table*}

As shown in Tab. \ref{table:b3lyp}, both for PEN and PFP, all known dipole-allowed transitions between singlets or between triplets are reproduced within deviations typical for TD-DFT calculations on similar systems.
In PEN, the lowest two calculated singlet transitions $S_1 \leftarrow S_0$ and $S_3 \leftarrow S_0$ agree within 0.05 eV with previous TD-DFT calculations using the same functional in a different basis set \cite{Grimme_2003_C_3_292}, but they underestimate the experimental transitions observed in the gas phase or in rare gas matrices~\cite{Amirav_1981_JoPC_85_309,Halasinski_2000_JoPCA_104_7484} by up to 0.5 eV.
Due to a significant gas-to-solvent shift, the difference between the calculated lowest transitions $S_1 \leftarrow S_0$ and the observed values is reduced to about 0.2 eV \cite{Hinderhofer_2007_Tjocp_127_194705}. 
Both in PEN and in PFP, the transition $S_1 \leftarrow S_0$ (peak $A_1$ and $B_1$, respectively) is dominated by a highest occupied molecular orbital/lowest unoccupied molecular orbital (HOMO/LUMO) excitation, so that these designations may be used synonymously.

For PEN, the very strong transition $S_6 \leftarrow S_0$ polarized along the long axis of the molecule is obtained in the TD-DFT calculation close to the value of 4.40 eV observed in a rare gas matrix \cite{Halasinski_2000_JoPCA_104_7484}. 
In PFP, the respective transition $S_6 \leftarrow S_0$ is dipole-forbidden.

In both molecules, the calculated lowest triplet state $T_1$ occurs below half of the lowest singlet transition energy, so that a non-radiative decay $S_1 + S_0 \rightarrow T_1 + T_1$ becomes energetically allowed. 
Using pump-probe spectroscopy for PEN, it was demonstrated that this singlet fission mechanism occurs on a time scale of about 80 fs \cite{Jundt_1995_CPL_241_84}, and the microscopic details of this process were analyzed with multi-reference perturbation theory \cite{Zimmerman_2010_NC_2_648}.
In PEN, the calculated triplet energy of 0.99 eV is in reasonable agreement with the experimental value of 0.85 eV. 
For PFP, we are not aware of any previous experimental or theoretical assignment of the lowest triplet energy.

The very weak transition $T_2 \leftarrow T_1$ in PEN has been observed with pump-probe spectroscopy in the region 1.2 to 1.4 eV as a transient absorption process from a long-living excited state \cite{Thorsmoelle_2009_PRL_102_17401}. 
As the ratio between the intensities of our observed PL bands $A_3$ and $A_4$ in PEN is temperature-independent, it is natural to assign both features to a vibronic progression with a fundamental transition $A_3$ at 1.52 eV, above all transient absorption features at lower energies assigned to the lowest triplet $T_1$. 
As the respective transition $T_2 \leftarrow T_1$ should show a small Stokes shift, only PL bands {\it slightly below} the highest observed absorption structure at 1.4 eV~\cite{Broch_2011_PRB_83_245307} might be assigned to this transition $T_2 \rightarrow T_1$. 
Therefore, we conclude that this recombination mechanism between the lowest two triplet states cannot contribute to our lowest fundamental PL band $A_3$ observed at 1.52 eV.

The rather strong transition $T_4 \leftarrow T_1$ has been detected both in solution \cite{Hellner_1972_JCSFT2_68_1928} and in polycrystalline films \cite{Jundt_1995_CPL_241_84}. 
In the latter case, it was suggested that this strong transition between triplets is approximately resonant with the lowest transition between singlets, $S_1 \leftarrow S_0$. 
Via the interaction between the transition dipoles involved, the reaction $S_0 + T_4 \rightleftharpoons S_1 + T_1$ (in our notation) might become a source for higher triplet states \cite{Jundt_1995_CPL_241_84}. 
As far as the observed PL spectra of PEN and PFP are concerned, a dipole-allowed recombination process $T_4 \rightarrow T_1$ between triplets might indeed contribute to the observed PL features above 2 eV. 
However, as the two highest PL transitions $A_5$ and $A_6$ are close to observed charge transfer transitions at 2.12 eV and 2.27 eV \cite{Sebastian_1981_CP_61_125}, we think that the respective recombination process involving adjacent cationic and anionic molecules provides a more natural assignment.
From calculated lattice sums of the interactions between the transition dipoles, it is evident that the excitonic dispersion possesses a minimum for vanishing wave vector ${\bf k}={\bf 0}$ at the $\Gamma$ point of the Brillouin zone \cite{Schlosser_1980_CP_49_181}. Therefore, the optically created Frenkel excitons thermalize around the dispersion minimum at ${\bf k}={\bf 0}$ from where they can recombine radiatively under momentum conservation.
Occasionally, excitons can scatter from Bloch waves into pairs of photocarriers. The resulting electrons and holes diffuse across the crystal before eventually localizing on a pair of adjacent molecules from where they can recombine radiatively. Even though the respective charge transfer transitions at 2.12 eV and 2.27 eV are high above the fundamental $E_{00}$ of the $S_1 \rightarrow S_0$ at $\Gamma$, scattering from the localized electron-hole pair into a Bloch wave at much lower energy seems to be inhibited by a high energy barrier.

To summarize this analysis, the observed PL bands $A_1$ in PEN at 1.83 eV and $B_1$ in PFP at $1.72$ eV are assigned to the fundamental of the $S_1 \rightarrow S_0$ recombination process at the $\Gamma$ point of the Brillouin zone. 
Moreover, in PFP, the satellites $B_2$ and $B_3$ can be related to a vibronic progression starting with $B_1$. The high energy transitions $A_5$ and $A_6$ in PEN are assigned to CT transitions between neighboring molecules observed in absorption at similar energies \cite{Sebastian_1981_CP_61_125}.
The low energy transition $A_2$ has been related to a self-trapped exciton (STE) state~\cite{Aoki-Matsumoto_2001_IjompB_15_3753}, and $A_3$ and $A_4$ represent a vibronic progression with the fundamental transition $A_3$.
  
\section{PEN:PFP blends}

In this section we present PL spectra of the coevaporated PEN:PFP thin films and discuss them in comparison to the pure samples. 

\subsection{1:1, 1:2, and 4:1 PEN:PFP mixed samples}
\begin{figure}
  \includegraphics[width=8cm]{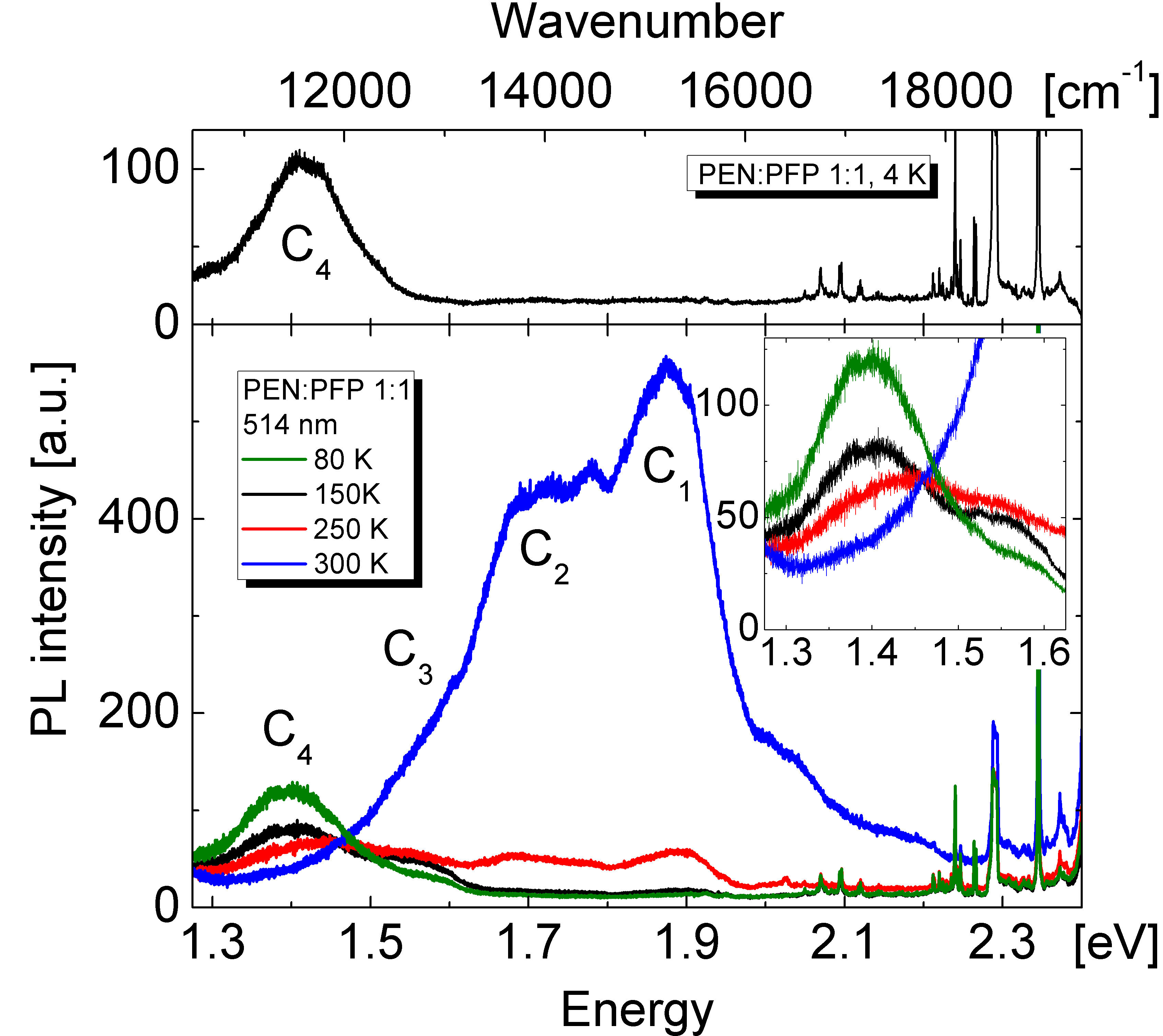}
  \caption{\label{Mix_1-1} PL spectra of a 1:1 mixed PEN:PFP thin film at different temperatures, excited at 514~nm (2.41~eV). Bottom: While the PL in the region above 1.6~eV decreases with decreasing temperature, a peak at 1.4~eV rises. Top: At 4~K only one peak at 1.4~eV remains, all other peaks vanish. Sharp lines appearing at higher energies are due to Raman scattering.}
\end{figure}

The existence of a pronounced radiative transition near 1.4~eV, which does not occur in the pure samples, is common to the PL of all blends. 
As will be shown below, it relates to a transition at 1.6~eV observed in absorption spectra of similar samples.~\cite{Broch_2011_PRB_83_245307} For the 1:1 mixture at 4~K it is the only radiative transition in the spectral range we investigated (Fig.~\ref{Mix_1-1}, top panel).

We classify the PL of the 1:1 blend into 4 bands 
($C_1$-$C_4$), see Fig.~\ref{Mix_1-1} (bottom).
At RT, we observe a dominant region of merged peaks ranging from approximately 1.5~eV to 2~eV. 
It shows a maximum at 1.88~eV ($C_1$) accompanied by a band at 1.7~eV ($C_2$) and a shoulder at 1.57~eV ($C_3$). 
The intensity of the merged bands $C_1$ and $C_2$ diminishes drastically at lower temperature and, interestingly, at approximately 150~K they disappear completely. 
Band $C_3$ at 1.57~eV decreases gradually towards lower temperature. 
We observe a new peak $C_4$ at 1.4~eV, continuously rising with decreasing temperature.

\begin{figure}
  \includegraphics[width=8cm]{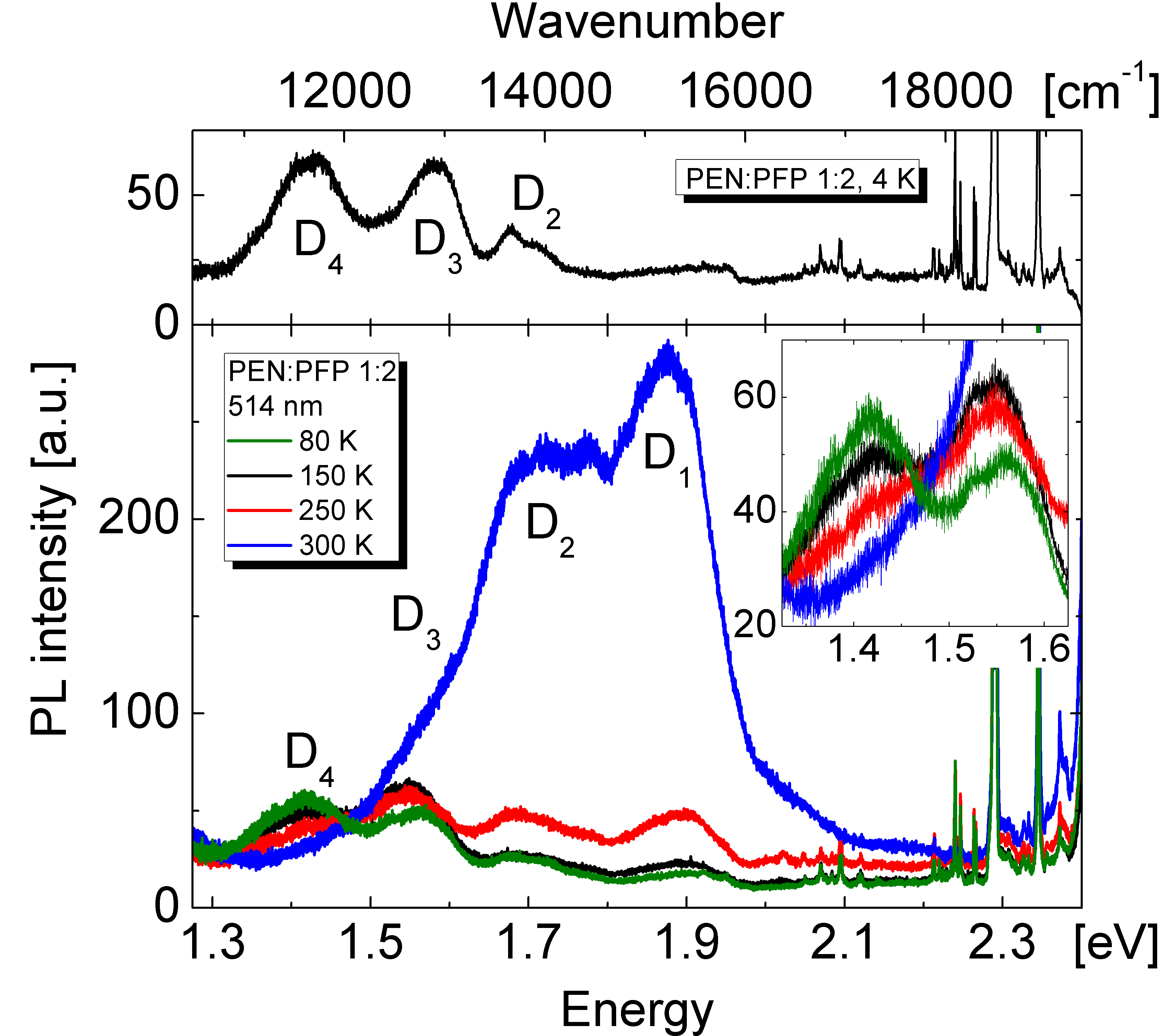}
  \caption{\label{Mix_1-2} PL spectra of a 1:2 mixed PEN:PFP thin film at different temperatures, excited at 514~nm (2.41~eV). Bottom: While the PL in the region above 1.6~eV decreases with decreasing temperature, a peak at 1.4~eV rises. Top: In contrast to the 1:1 blend, there remain three bands at 4~K, where $D_2$ remains near the HOMO/LUMO transition of PFP, and $D_4$ corresponds to $C_4$ in the 1:1 blend. For further details, see text. Sharp lines appearing at higher energies are due to Raman scattering.}
\end{figure}

Similar to the 1:1 blend, the 1:2 mixture (Fig.~\ref{Mix_1-2}) with more PFP exhibits merged bands at RT ranging from approximately 1.5~eV to 2~eV.
Though with slight modifications, it shows shoulders and peaks in similar regions as the 1:1 mixture, which we denote as $D_1$ (1.88~eV), $D_2$ (1.7~eV) and $D_3$ (1.56~eV), corresponding to $C_1$, $C_2$, and $C_3$, respectively.
The shoulder $D_2$ in the region around 1.7~eV is slightly more pronounced compared to $D_1$ than the corresponding shoulder $C_2$ in the 1:1 mixture.

The bands $D_1$ and $D_2$ decrease with temperature but a peak at approximately 1.7~eV ($D_2$) still remains at 4~K, in contrast to the 1:1 blend. 
The shoulder denoted as $D_3$ decreases with temperature, reaches a minimum at around 100~K, is blueshifted by $\sim$20~meV and rises again towards lower temperatures, which is also different from the 1:1 blend.
Below 300~K, a band $D_4$ at 1.4~eV (corresponding to $C_4$) increases continuously towards lower temperatures.

\begin{figure}
  \includegraphics[width=8cm]{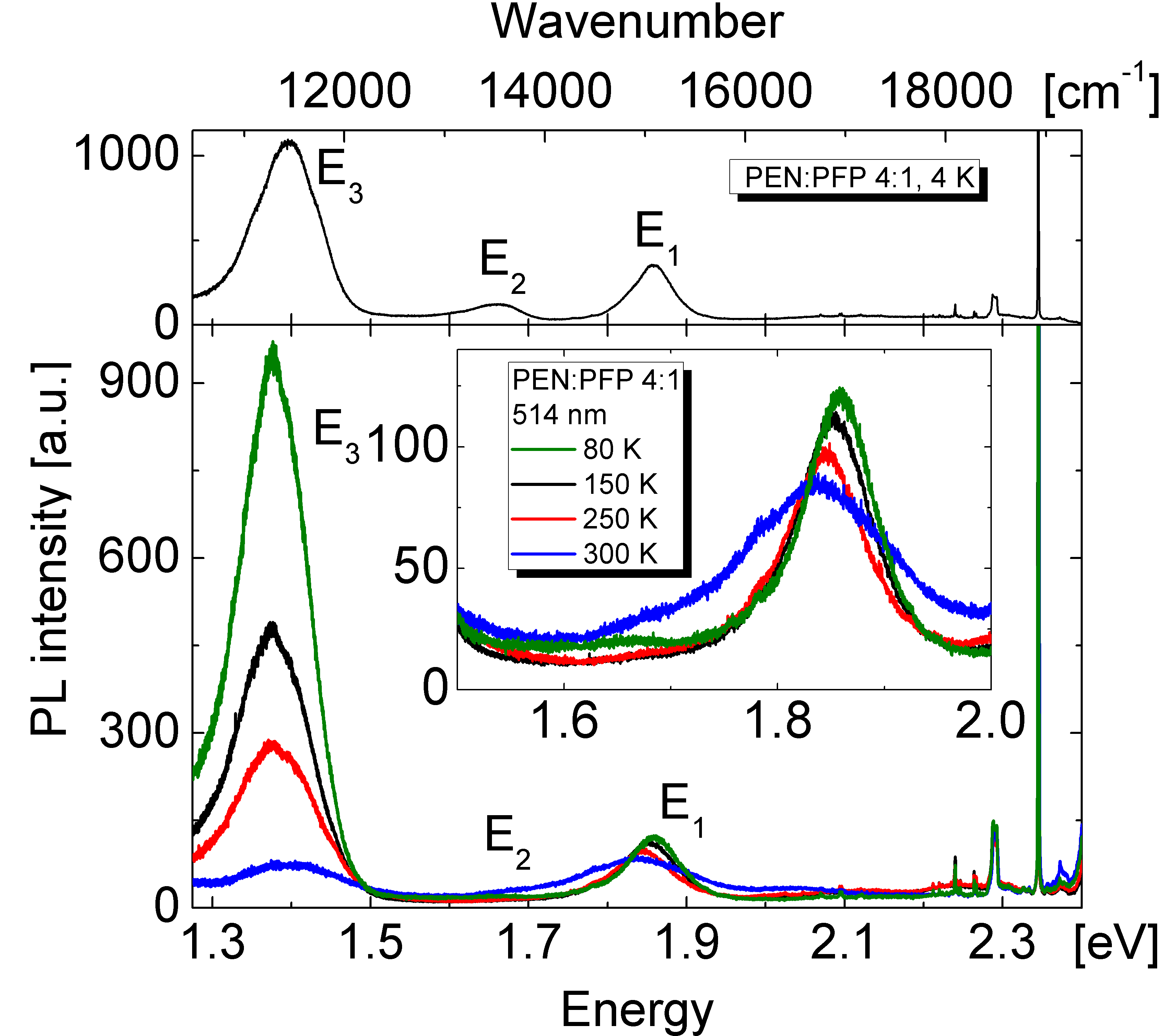}
  \caption{\label{Mix_4-1} PL spectra of a 4:1 mixed PEN:PFP thin film at different temperatures, excited at 514~nm (2.41~eV). Bottom: At RT, there are two pronounced peaks at 1.83~eV ($E_1$) and 1.4~eV ($E_3$). Towards low temperatures, band $E_3$ rises drastically, $E_2$ appears below 100~K. Top: At 4~K, there remain three bands: While $E_1$ and $E_2$ resemble PEN peaks $A_1$ and $A_2$, respectively, the third peak $E_3$ has the same energy as $C_4$ in the 1:1 blend, but is more pronounced. For further details see text. Sharp lines appearing at higher energies are due to Raman scattering.}
\end{figure}

For the third blend (Fig.~\ref{Mix_4-1}) with the PEN:PFP mixing ratio 4:1, three PL features $E_1$ to $E_3$ can be identified at 1.83~eV, 1.66~eV, and 1.41~eV, respectively. 
The bands of this mixture are separated and can be well resolved, $E_1$ and $E_3$ are both pronounced at RT.
$E_1$ has generally the shape and position of band $A_1$ of PEN, however it is broader and shows a shoulder around 1.7~eV (see inset). Towards lower temperatures $E_1$ suffers a blueshift that is comparable to that of $A_1$ in PEN. 
$E_3$ increases continuously upon cooling down and becomes the dominant peak at low temperatures. 
We can observe a redshift of the peak by approximately 20~meV at low temperatures.

Below 100~K we observe a band $E_2$, exhibiting an asymmetry resembling the shape of band $A_2$ in PEN. 
The ratio of the amplitude of $E_2$ and $E_1$ is equal to the ratio of $A_2$ to $A_1$ of PEN.
Upon excitation with the 2.54~eV line of the Ar$^+$-laser, we observe two peaks at 2.12~eV and 2.29~eV (not shown), analogously to the pure PEN spectrum.
Tab.~\ref{tab:Mix} gives an overview of the peaks observed for the three different PEN:PFP blends. 

\begin{table}[!ht]
  \centering
  \begin{tabular}{ll|l}
    \hline
    \hline
    \multicolumn{3}{c}{PEN:PFP 1:1} 	\\
    \hline
    $C_1$ & 1.88~eV& $S_1 \rightarrow S_0$ of PEN \\
    $C_2$ & 1.7~eV& $S_1 \rightarrow S_0$ of PFP \\
    $C_3$ & 1.57~eV& \\
    $C_4$ & 1.4~eV& Likely CT PEN $\leftrightarrow$ PFP \\
    \hline
    \multicolumn{3}{c}{PEN:PFP 1:2} 	\\
    \hline
    $D_1$ & 1.88~eV& $S_1 \rightarrow S_0$ of PEN \\
    $D_2$ & 1.7~eV& $S_1 \rightarrow S_0$ of PFP \\
    $D_3$ & 1.56-1.58~eV&  \\
    $D_4$ & 1.4~eV& Likely CT PEN $\leftrightarrow$ PFP \\
    \hline
    \multicolumn{3}{c}{PEN:PFP 4:1} 	\\
    \hline
    $E_1$ & 1.83-1.84~eV& $S_1 \rightarrow S_0$ of PEN  \\
    $E_2$ & 1.66~eV& STE of PEN \\
    $E_3$ & 1.41-1.39~eV& Likely CT PEN $\leftrightarrow$ PFP \\
    \hline
    \hline
  \end{tabular}
  \caption{\label{tab:Mix} List of peaks of the blended samples and their tentative assignment to transitions in relation to pure PEN and PFP. For further 	details, see text.}
\end{table}

\subsection{Discussion of blends.} \label{Sec:4_Discussion}

Obviously, the PL spectra of the mixtures differ strongly from the pure samples and cannot be regarded as simple superpositions of the PEN and PFP spectra, neither at RT nor at low temperatures. 
Importantly, at 4~K the 1:1 blend exhibits only one pronounced peak at around 1.4~eV ($C_4$) that does not occur either in pure PEN or PFP. 
Other peaks observable in the pure materials vanish completely, in particular in the region of the respective HOMO/LUMO transitions.
Altogether, we consider this as clear evidence for intermolecular coupling of PEN and PFP. 
Since a good theoretical understanding of molecular mixed thin films is missing to a large extent, a unique assignment of PL peaks is difficult.

\begin{figure}[htbp]
  \includegraphics[width=8cm]{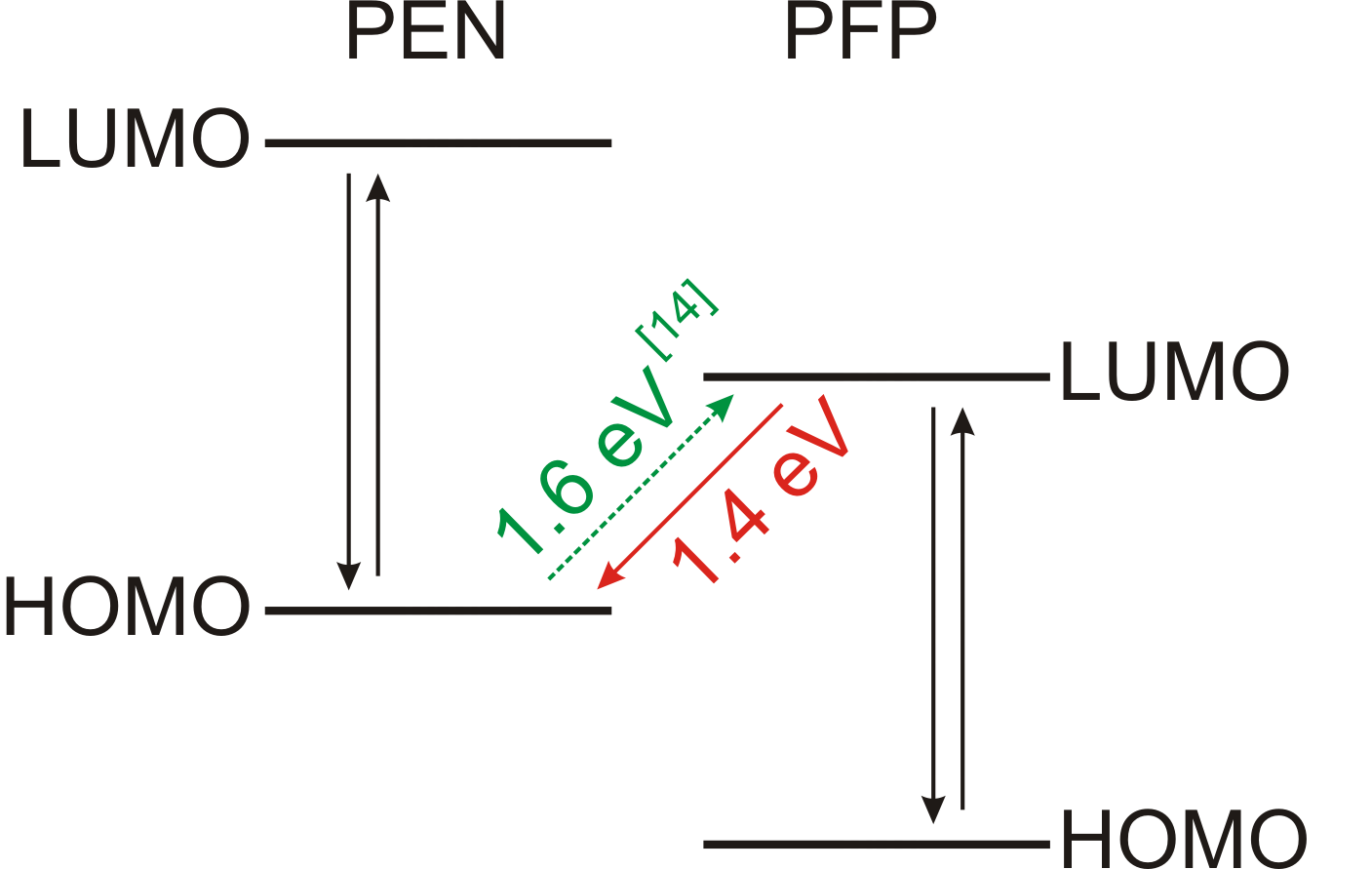}
  \caption{\label{CT} Simplified summary of optical transitions in PEN:PFP blends. A possible explanation for the origin of the peak at 1.4~eV (red) observed with PL and a corresponding transition reported from absorption measurements~\cite{Broch_2011_PRB_83_245307} (green dotted line) of a mixed PEN:PFP thin film could be charge transfer (CT). The values stem from thin films of 20~nm thickness. Note that the values relate to optical transitions; energy level distances would have to account for the exciton binding energy.}
\end{figure}

Concerning PEN:PFP blended thin films, in particular for equimolar mixtures, X-ray diffraction experiments~\cite{Hinderhofer_2011_JoCP_134_104702} indicate good intermixing of both materials on a molecular level. 
The disappearance of the $S_1 \rightarrow S_0$ transition of PEN and PFP at low temperatures in the blends raises interesting questions. 
Scaling with the mixing ratio, the suppression of the respective bands is most pronounced for the 1:1 mixture.
For this reason, we relate it to intermolecular coupling effects between PEN and PFP. 
In the mixtures, excitonic states which would recombine radiatively in the pure phases may decay non-radiatively or transform into the state emitting at 1.4 eV, presumably a CT state delocalized over a hetero-interface as visualized in Fig.~\ref{CT}.
A detailed analysis of the underlying mechanisms would require extensive calculations beyond the scope of the present paper.
The existence of peaks in the region of the HOMO/LUMO transition of PEN and PFP, which we observe in the non-1:1 blended thin films, is consistent with X-ray diffraction results, where the formation of mixed PEN:PFP phases and pure PEN or PFP phases was reported.~\cite{Hinderhofer_2011_JoCP_134_104702}
Hence, the transition at 1.4~eV is likely to originate from the intermixed phases forming also in the non-1:1 blended thin films, whereas 
transitions resembling pure PEN ($C_1$, $D_1$, $E_1$, $E_2$) or PFP ($C_2$, $D_2$) might be assigned to the respective pure phases in the mixtures.
It appears reasonable that $E_1$ in the sample with a higher PEN ratio stems from the PEN $S_1 \rightarrow S_0$ transition, whereas peak $D_2$ of the sample with more PFP corresponds to the ($\sim$30~meV blueshifted) PFP $S_1 \rightarrow S_0$ transition. 
The origin of a band in the mixtures at around 1.57~eV ($C_3$ and $D_3$) remains unclear.
Similar to CT transitions in pure PEN involving different geometries of the molecule pair~\cite{Sebastian_1981_CP_61_125}, it could be due to a CT transition between more distant molecules or due to a trap state, significantly below the $S_1 \rightarrow S_0$ transition energies of both pure phases of PEN and PFP.

A reasonable explanation for the transition at 1.4~eV could be that this band is related to a CT between the two components, particularly a CT between PEN and PFP is generally promoted by good molecular intermixing as found by X-ray diffraction~\cite{Hinderhofer_2011_JoCP_134_104702,Salzmann_2008_L_24_7294}.
While the suppression of PEN and PFP bands in their respective region of HOMO/LUMO transitions is maximized for the equimolar mixture, we find that the intensity of the transition at 1.4~eV is favored by a higher PEN ratio.
The only peak we observe in this region in the pure materials occurs for PEN ($A_4$). 
However, transition $A_4$ in PEN, assigned to a vibronic progression of $A_3$, cannot explain the transition at 1.4 eV in the blends 
because $A_3$ does not occur in any of the mixtures.
Recombination processes $T_2 \rightarrow T_1$ are expected to be still less likely than in pure PEN films, because singlet fission should be suppressed by more efficient charge separation processes between PEN and PFP. This phenomenon was demonstrated already for PEN:C$_{60}$\cite{Thorsmoelle_2009_PRL_102_17401}, where singlet fission is suppressed, so that the transient absorption from the lowest triplet $T_1$ is reduced accordingly.
  
Thus, we conclude that the band at 1.4 eV in the mixture cannot stem from any of the two single components alone. 
Instead, it represents a new transition that might be due to a CT between the components, i.e. related to a similar transition observed recently in absorption spectra of PEN:PFP blends at 1.6 eV~\cite{Broch_2011_PRB_83_245307}.
As this absorption feature cannot be interpreted in terms of the known singlet transitions in the pure materials summarized in Tab. \ref{table:b3lyp} and Sec. \ref{Sec:3c_Transitions}, a weak charge transfer transition from the HOMO in PEN to the LUMO in PFP becomes the most likely assignment.
Considering the relative alignment of the HOMO/LUMO levels of PEN and PFP as they result from photoelectron spectroscopy measurements~\cite{Salzmann_2008_JotACS_130_12870} and DFT calculations~\cite{Sakamoto_2006_MCaLC_444_225,Medina_2007_JoCP_126_111101},
Fig.~\ref{CT} sketches a possible energy diagram involving a CT.

\begin{table}[htbp]
	\begin{tabular}{c|c|c|c}
		\hline
		\hline
		\multicolumn{4}{c}{mixed PEN:PFP} \\
		\hline
		CT transition	energy &	1.60 eV$^a$	&	1.40 eV$^b$ &	1.73 eV$^c$ \\
		\hline
		\hline
	\end{tabular}
	\caption{CT transition energies as obtained by $^a$ absorption~\cite{Broch_2011_PRB_83_245307} (LUMO (PFP) $\leftarrow$ HOMO (PEN)), $^b$ PL peak $C_4$, $D_4$, and $E_3$ (LUMO (PFP) $\rightarrow$ HOMO (PEN)), and $^c$ from HOMO positions deduced from UPS on mixed PEN:PFP films \cite{Salzmann_2008_JotACS_130_12870} together with DFT at the B3LYP/6-31G(d) level, see Supplementary Material for further details.}
	\label{table:CT}
\end{table}

\subsection{Assignment of CT transition between PEN and PFP}

\label{Sec:4B_Assignment}

In order to shed more light on the above data, DFT results for the energies of the frontier orbitals and for the ionization potential (IP) may be compared to published ultraviolet photoemission spectra (UPS) and optical transition energies. 
The IP is calculated as the difference between a neutral and a cationic molecule. 
In ultraviolet photoemission (UPS), multilayers of flat-lying PEN or PFP deposited on gold reveal HOMO binding energies and IPs of 5.45~eV or 6.20 - 6.40~eV below the respective vacuum level \cite{Koch_2007_AM_19_112}.
These values are about 0.4 - 0.5~eV below the DFT values for the ionization potential. Such changes in the absolute value of the IP can be related to the polarization of molecules surrounding the cationic site in the surface layer \cite{Tsiper_2003_PRB_68_85301}.
Nevertheless, the difference between the measured HOMO levels of 0.75 to 0.95~eV is well reproduced by the calculated value of 0.79~eV, indicating that polarization effects in pure PEN and pure PFP films composed of lying molecules induce similar energy shifts. 
The calculated binding energies of the HOMO levels of both compounds remain 1.33~eV below the calculated IP, a deficiency of DFT which can be related to the wrong asymptotics of the exchange-correlation potential.\cite{Leeuwen_1994_PRA_49_2421}
The difference between the calculated frontier orbitals is 2.21~eV for PEN and 2.03~eV for PFP, in each case about 0.2~eV above the average energy of the observed transition energy.\cite{Broch_2011_PRB_83_245307}

When pure PEN or PFP films are grown on SiO$_2$ like in the present work, the molecules adsorb close to upright. 
In this case, the opposite polarity of C-H and C-F bonds in the two compounds results in substantial changes of the UPS spectra, so that the difference between the HOMO positions increases to 1.9 eV.\cite{Salzmann_2008_JotACS_130_12870} 
However, for a mixed phase of standing molecules with mixing ratio 1:1, both compounds establish a common vacuum level so that the HOMO levels move towards each other, resulting in a much smaller splitting between $-6.6$~eV for PFP and $-6.3$~eV for PEN.\cite{Salzmann_2008_JotACS_130_12870}
When using these UPS values for the HOMO energies together with the difference between the frontier orbitals calculated with DFT, one can deduce LUMO energies for the molecules in the mixed phase. 
On this basis, the difference between the LUMO of PFP and the HOMO of PEN gives an estimate of 1.73~eV for the CT gap between the two compounds, or about 0.13~eV above the observed absorption band at 1.60~eV reported in Tab. \ref{table:CT} \cite{Broch_2011_PRB_83_245307}. 
This small difference indicates that energy corrections arising from the Coulomb interaction between the electron and the hole forming the CT state across the PEN-PFP interface remain somewhat smaller than the excitonic binding energy in the pure compounds.

Under the assumption that the excitonic binding energy in the pure compounds results in an average transition energy of 0.2~eV below the DFT gap energies, and that due to the Coulomb attraction between oppositely charged molecules, the above estimate for the CT state has to be reduced by 0.13~eV, we come to the assignment of the observed absorption bands visualized in Fig. \ref{CT}. 
Hence, together with a Stokes shift of 0.2~eV arising from reorganization energies of about 0.1~eV for either positively charged PEN or for negatively charged PFP, we can assign the observed PL band at 1.4 eV to the CT transition in the mixed phase.
The available experimental data are thus all nicely consistent with our interpretation.

\section{Summary and conclusion}

We have studied PL spectra of PEN, PFP, and PEN:PFP mixed films grown on SiO$_2$. For PFP, we find fewer radiative transitions than for PEN and identify a peak at 1.72~eV as the HOMO/LUMO transition exhibiting a vibronic progression.
In the mixed PEN:PFP system, we find evidence for coupling in equimolar PEN:PFP blends.~\cite{Broch_2011_PRB_83_245307} 
Based on experiments and theoretical analysis, we suggest that the transition at 1.4~eV originates from a CT state, i.e. similar to absorption measurements showing a corresponding feature at 1.6~eV.~\cite{Broch_2011_PRB_83_245307}
We believe that our experimental data along with broader theoretical interpretation help to understand the important issue of donor/acceptor coupling.

\section {acknowledgment}

We gratefully acknowledge financial support by the DFG, the DAAD, and the Studienstiftung des Deutschen Volkes.

%

\end{document}